\documentclass[twocolumn]{aastex63}
\usepackage{amsmath}

\usepackage{graphicx}
\usepackage{amsmath}
\usepackage{mathtools}
\usepackage{kantlipsum}
\usepackage{xparse}
\usepackage{xcolor}
\usepackage{enumerate}
\usepackage[version=3]{mhchem}
\usepackage{amsmath,array}
\usepackage[T1]{fontenc}
\usepackage{booktabs}

\usepackage{sidecap}
\usepackage{tabularx}
\usepackage{float}
\usepackage{xspace}
\usepackage{rotating}

\newcommand\eureka{{\texttt{Eureka!}}\xspace}
\renewcommand\micron{\textmu m}
\newcommand\microns{\micron}
\newcommand{\SteinrueckEtAlT}{Steinrueck et al. (in review)}

\received{Received}
\revised{Revised}
\accepted{Accepted}
\submitjournal{ApJ}

\graphicspath{{./}{figures/}}

\begin{document}

\title{Clouds and Hazes in GJ 1214b’s Metal-Rich Atmosphere}

\correspondingauthor{Isaac Malsky}
\email{isaac.n.malsky@jpl.nasa.gov}

\author[0000-0003-0217-3880]{Isaac Malsky}
\affil{Department of Astronomy and Astrophysics, University of Michigan, Ann Arbor, MI, 48109, USA}
\affil{Jet Propulsion Laboratory, California Institute of Technology, Pasadena, CA 91109, USA}

\author[0000-0003-3963-9672]{Emily Rauscher}
\affil{Department of Astronomy and Astrophysics, University of Michigan, Ann Arbor, MI, 48109, USA}

\author[0000-0002-7352-7941]{Kevin Stevenson}
\affil{Johns Hopkins Applied Physics Laboratory, Laurel, MD, 20723, USA}

\author[0000-0002-2454-768X]{Arjun B.\ Savel}
\affiliation{Department of Astronomy, University of Maryland, College Park, MD, USA}
\affiliation{Center for Computational Astrophysics, Flatiron Institute, New York, NY, USA}

\author[0000-0001-8342-1895]{Maria E. Steinrueck}
\altaffiliation{51 Pegasi b fellow}
\affil{Department of Astronomy \& Astrophysics, University of Chicago, Chicago, IL USA}
\affiliation{Max-Planck-Institut f\"ur Astronomie, Heidelberg, Germany}

\author[0000-0002-8518-9601]{Peter Gao}
\affiliation{Earth \& Planets Laboratory, Carnegie Institution for Science, Washington, DC, USA}

\author[0000-0002-1337-9051]{Eliza M.-R. Kempton}
\affiliation{Department of Astronomy, University of Maryland, College Park, MD, USA}

\author[0000-0001-8206-2165]{Michael T.\ Roman}
\affiliation{School of Physics and Astronomy, University of Leicester, Leicester, UK}

\author[0000-0003-4733-6532]{Jacob L.\ Bean}
\affil{Department of Astronomy \& Astrophysics, University of Chicago, Chicago, IL USA}

\author[0000-0002-0659-1783]{Michael Zhang}
\altaffiliation{51 Pegasi b fellow}
\affil{Department of Astronomy \& Astrophysics, University of Chicago, Chicago, IL USA}

\author[0000-0001-9521-6258]{Vivien Parmentier}
\affiliation{Department of Physics, University of Oxford, Oxford, UK}

\author[0000-0002-4487-5533]{Anjali A.\ A.\ Piette}
\affiliation{School of Physics and Astronomy, University of Birmingham, Edgbaston, Birmingham B15 2TT, UK}

\author[0000-0003-3759-9080]{Tiffany Kataria}
\affil{Jet Propulsion Laboratory, California Institute of Technology, Pasadena, CA 91109, USA}

\begin{abstract}
The sub-Neptune GJ 1214b has an infamously flat transmission spectrum, likely due to thick aerosols in its atmosphere. A recent JWST MIRI spectroscopic phase curve of GJ 1214 b added to this picture, suggesting a highly reflective and metal-rich atmosphere. Using a 3D General Circulation Model with both photochemical hazes and condensate clouds, we characterize how different aerosol types affect the atmospheric structure of GJ 1214 b and manifest in its spectroscopic phase curve. Additionally, we reanalyze the original GJ 1214 b JWST phase curve. The reanalysis shows a hotter nightside, similar dayside temperature, and a lower, but still elevated, Bond albedo (0.42 $\pm$ 0.11) than the original results. We find that a scenario with both clouds and hazes is most consistent with the JWST phase curve. Reflective clouds or hazes are needed to explain the large Bond albedo, and hazes or a super-solar metallicity help account for the several hundred Kelvin day-night temperature difference measured by the phase curve.
\end{abstract}

\section{Introduction}\label{sec:Introduction}
GJ 1214 b is an archetypal sub-Neptune \citep{Charbonneau2009}, initially theorized to either be a rocky core with an atmosphere of H/He or a water-rich world with a heavier mean molecular weight atmosphere \citep{Leger2004, Bean2011, Ricci2010, Nettelmann2011, Menou2012, Valencia2013}. The origin and evolution of sub-Neptunes is an open question in exoplanet astronomy \citep[e.g.,][]{Bean2021}, made especially interesting because of their abundance \citep{Zhu2021}, and because this class of planet has no Solar System analogue. Measuring the atmospheric abundances of volatiles will break degeneracies between drift and migration models, as the pebbles in the drift model are expected to lose volatiles \citep{Bean2021}.

Aerosols (clouds and hazes) are pervasive in exoplanets, and have frustrated efforts to accurately constrain the physical and chemical structure of exoplanets. Photochemical hazes are anticipated on planets with equilibrium temperatures below 1000 Kelvin \citep[e.g.,][]{Gao2020, Yu2021}, while condensate clouds are expected for a wide range of temperatures \citep[e.g.,][]{Heng2013, Kreidberg2014, Sing2016}. Aerosols can dramatically alter a planet's energy balance, influence the atmospheric structure, lead to flat transmission spectra, and shape emission spectra (for detailed reviews, see \citealt{Marley2013}, \citealt{Helling2019}, and \citealt{Gao2021a}). Additionally, the formation of aerosols can impact bulk atmospheric abundances. For example, the sequestration of oxygen in clouds influences atmospheric C/O ratios \citep{Lee2016, Helling2023}.

Untangling the effects of aerosols on atmospheric structure and observables is complicated by uncertainties in formation mechanisms, optical properties, and vertical and horizontal mixing \citep{Helling2019}. A variety of cloud particle sizes and compositions, including silicates, metal oxides, salts, and sulfides are possible. The optical properties and formation rates of hazes are also not precisely characterized. Titan-like tholins, hydrocarbon soot, or sulfur hazes are all \textit{a priori} possible \citep{Gao2021a}. By comparing self-consistent numerical models that span a range of aerosol parameterizations with high-fidelity observations over a wide wavelength range, it is possible to constrain atmospheric metallicities and aerosol distributions.

Previous observational studies of GJ 1214 b have revealed a flat, featureless transmission spectrum, indicating that GJ 1214b has an atmosphere blanketed with aerosols, and likely also has a high mean molecular weight \citep{Bean2010, Crossfield2011, Narita2013, Morley2013, Kreidberg2014}. Recently, \cite{Kempton2023} observed the first thermal emission phase curve of GJ 1214 b with JWST, providing a new avenue for insight into the atmospheric composition of this planet. This observation showed dayside and nightside temperatures of 553 $\pm$ 9 K and 437 $\pm$ 19 K, respectively, and a Bond albedo of 0.51 $\pm$ 0.06 --- indicative of reflective upper atmosphere aerosols and a super-solar metallicity. Furthermore, the spectroscopic observation showed strong (moderate) evidence of absorption features on the nightside (dayside). \cite{Kempton2023} included 3D General Circulation Models (GCMs) with various haze assumptions and metallicities, and found that the best match to 3D models had super-solar metallicity compositions ($>$100x) and thick reflective hazes. However, no combination of atmospheric metallicity and photochemical haze prescription fit the JWST data, motivating the existence of missing physics or chemistry in the GCMs. In this work we focus on the impact that clouds, in addition to hazes, have on sculpting the global thermal emission from GJ 1214b.

Three-dimensional simulations of exoplanets provide a deeper understanding of their underlying atmospheric structure and dynamics, as well as help interpret observational data \citep[e.g.,][]{Showman2009, dobbsdixon2010, Rauscher2010, Heng2013, Mayne2014, Mendon2016, Lee2016}. Previous works have used GCMs to predict the atmospheric circulation pattern of GJ 1214 b and study how it is influenced by metallicity, clouds, and model convergence times \citep[e.g.,][]{Menou2012, Kataria2014,Charnay2015a, Wang2020b, Christie2022}. Briefly, \cite{Menou2012} found super-rotating zonal winds to be robust across solar, super-solar, and water compositions—but showed variations in the degree of eastward heat advection between models. \cite{Kataria2014} was the first to model GJ 1214 b with multiwavelength radiative transfer and found high mean molecular weight atmospheres produced larger day-night and equator-to-pole temperature differences. \cite{Wang2020b} demonstrated changes in circulation patterns with longer simulation run-times, due to the deep atmosphere's large radiative timescale. \cite{Charnay2015a} and \cite{Christie2022} modeled GJ 1214 b with radiatively coupled clouds, and showed that clouds increase planetary albedo. In addition, \cite{Christie2022} showed that clouds' impact on thermal structure and, consequently, on phase curves, is influenced by the sedimentation factor. Shallow clouds have relatively little impact on the physical structure of GJ 1214 b. However, no study so far includes both photochemical hazes and clouds.

In this work, we investigate the differences in atmospheric structure of GJ 1214 b for a range of metallicities, hazes, and cloud prescriptions, in order to interpret the recent phase curve results from \cite{Kempton2023}. In particular, we present the first sub-Neptune models with both condensate clouds and photochemical hazes. Both classes of aerosols have been hypothesized in order to explain the composition and transmission spectra of GJ 1214 b \citep[e.g.,][]{millerricci2012,Morley2013,Morley2015, Charnay2015a, gao2018b, ohno2018, ohno2020b, adams2019, kawashima2019a, lavvas2019, Christie2022}.

This work is organized as follows. In \S~\ref{sec:methods}, we describe the GCM used for this work and the grid of models that we run. We outline the additions to the GCM necessary to simulate GJ 1214 b with both radiatively active clouds and hazes. This marks the first time that both phenomena have been self-consistently included in a published exoplanet GCM. In \S~\ref{sec:results}, we present how these different model parameterizations affect the resulting planetary structure in terms of pressure-temperature profiles, winds, cloud condensation, and global energy balance. Then, we generate simulated emission spectra and phase curves and compare the results to a reanalysis of the phase curve from \cite{Kempton2023} in \S~\ref{sec:JWST_compare}. In \S~\ref{sec:conclusions}, we summarize our main conclusions regarding the composition and atmospheric structure of this planet. The methodology of the phase curve reanalysis is presented in the Appendix.

\section{Methodology}\label{sec:methods}
The GCM used for this work, the RM-GCM, has been adapted from its initial creation as a Newtonian relaxation model for Earth meteorology \citep{Hoskins1975}. It has been expanded upon to simulate hot Jupiters and sub-Neptunes \citep{Menou2009, Rauscher2010, Rauscher2012, Rauscher2013, Roman2017, Roman2019, May2020, Roman2021, Beltz2022, Malsky2024}. These updates include capabilities to model double gray radiative transfer, radiatively active temperature-dependent clouds, and clouds with pressure dependent scattering and absorption. Here, we make two additions to the GCM. First, we include the effects of radiatively active photochemical hazes. Second, we expand upon the cloud modeling within the GCM and now include cloud condensation curves for higher atmospheric metallicities.

\subsection{GCM Framework}\label{subsec:GCMS}
The planet and star parameters used in this work are shown in Table \ref{tab:planet_parameters}. For all of our simulations, we assume that the planet synchronously rotates with an orbital period that matches its rotational period. We model the atmosphere from 10$^{-5}$ to 10$^{2}$ bar, divided into 50 layers with logarithmic pressure spacing. We implement a T31 horizontal spectral resolution, translating to a grid of 48 latitude by 96 longitude points. Each simulation was run for 1000 model days, at 4800 time-steps per day (approximately 10 Earth days of simulation time on a Intel Xeon Gold 6154 CPU). The initial pressure-temperature profile was calculated using the analytical solution from \cite{Guillot2010}, and an initial heat redistribution efficiency factor of 0.25 \citep{Burrows2003}. For all models, we use the double gray method to calculate gas opacities. Although \cite{Malsky2024} recently added picket fence radiative transfer to the RM-GCM, this method has not yet been expanded for non-G type stars \citep{parmentier2015} and the double gray method has been previously used to model GJ 1214 b \citep[e.g.,][]{Menou2012}.

We calculate the radiative transfer in two distinct channels, one to represent the thermal emission of the planet, and one to represent the starlight incident on the planet. For the radiative transfer scheme, we use the two-stream approximation for inhomogeneous multiple-scattering atmospheres \citep{Toon1989}. The absorption coefficients that we use for the double gray radiative transfer are shown in Table \ref{tab:planet_parameters}. For the 1x and 30x solar metallicity cases, we use the same thermal and starlight absorption coefficients as in \cite{Menou2012}. \cite{Menou2012} calculated these values to match 1D temperature–pressure profiles of GJ 1214 b from \cite{MillerRicci2010}. The \cite{Menou2012} models do not include a 100x solar metallicity case, and for this model we use HELIOS \citep{Malik2017, Malik2019} to calculate 1D temperature pressure profiles in radiative-convective equilibrium. We chose the thermal and starlight absorption coefficients such that the analytic temperature-pressure profiles \citep{Guillot2010} match these profiles as closely as possible.

\begin{deluxetable*}{lll}
{\tablehead{\multicolumn{3}{c}{System Parameters}}}
\startdata
\multicolumn{1}{c}{Parameter} & \multicolumn{1}{l}{Value} & \multicolumn{1}{l}{Units} \\
\hline
\hline
Semi-major Axis                                        & 0.0149                 & au                \\
Stellar Effective Temperature                          & 3250                   & K                 \\
Stellar Radius                                         & 0.215                  & R$_\odot$         \\
Planet Gravitational Acceleration                             & 10.65                  & m s$^{-2}$        \\
Planet Radius                                          & 1.747$\times$10$^7$    & m                \\
Intrinsic Temperature (T$_{int}$)                       & 100                    & K                \\
Planetary Rotation Rate                                & 4.615$\times$10$^{-5}$ & radians s$^{-1}$  \\
\hline
\hline 
\multicolumn{3}{c}{Solar Metallicity Specific Parameters} \\
\hline
\hline
Ratio of gas constant to heat capacity  ($R/c_p$)       & 0.286	                 & dimensionless   \\
Specific gas constant $R$                               & 3574	                 & J kg$^{-1}$ K$^{-1}$   \\
Thermal absorption coefficient $\kappa_{\mathrm{IR}}$   & 2.00$\times$10$^{-3}$  & m$^2$ kg$^{-1}$   \\
Optical absorption coefficient $\kappa_{\mathrm{VIS}}$  & 8.00$\times$10$^{-5}$  & m$^2$ kg$^{-1}$   \\
Infrared photosphere ($\tau$=2/3, cloud free)           & 36                     & mbar          \\
\hline
\hline
\multicolumn{3}{c}{30x Solar Metallicity Specific Parameters} \\
\hline
\hline
Ratio of gas constant to heat capacity  ($R/c_p$)       & 0.286	                 & dimensionless   \\
Specific gas constant $R$                               & 3299	                 & J kg$^{-1}$ K$^{-1}$   \\
Thermal absorption coefficient $\kappa_{\mathrm{IR}}$   & 1.00$\times$10$^{-2}$  & m$^2$ kg$^{-1}$   \\
Optical absorption coefficient $\kappa_{\mathrm{VIS}}$  & 4.00$\times$10$^{-4}$  & m$^2$ kg$^{-1}$   \\
Infrared photosphere ($\tau$=2/3, cloud free)           &  7                     & mbar            \\   
\hline
\hline
\multicolumn{3}{c}{100x Solar Metallicity Specific Parameters} \\
\hline
\hline
Ratio of gas constant to heat capacity  ($R/c_p$)       & 0.273                  & dimensionless   \\
Specific gas constant $R$                               & 1917	                 & J kg$^{-1}$ K$^{-1}$   \\
Thermal absorption coefficient $\kappa_{\mathrm{IR}}$   & 6.00$\times$10$^{-2}$  & m$^2$ kg$^{-1}$   \\
Optical absorption coefficient $\kappa_{\mathrm{VIS}}$  & 1.8$\times$10$^{-3}$   & m$^2$ kg$^{-1}$   \\
Infrared photosphere ($\tau$=2/3, cloud free)           & 1                      & mbar            \\   
\enddata
\label{tab:planet_parameters}
\caption{The parameters used for our models of GJ 1214 b.}
\end{deluxetable*}

\subsection{Cloud Condensation Curves}\label{sec:cloud_condensation_curves}
In this work, we also add the capability of simulating higher metallicity clouds to the RM-GCM. We follow the same methodology as presented in \cite{Mbarek2016}, and generate condensation curves for 1x, 30x, and 100x solar metallicity. The condensation curve determines the pressure-temperature relationship for which the vapor pressure of a cloud species in gas form equals to its saturation vapor pressure \citep{Morley2013}. Higher metallicities increase vapor pressures, pushing the condensation curve to higher temperatures. To calculate the pressure-temperature condensation curves at super solar metallicities, we adjust the initial \cite{Lodders2003} abundances of all species other than H and He by the desired metallicity enhancement factor. Next, we calculate the chemical equilibrium values with rain-out and find the temperature where each of the limiting elemental species condenses out.

Following the methodology in \cite{Roman2017} and \cite{Roman2019}, we include radiatively active clouds within the RM-GCM. Using data obtained from \cite{Burrows1999}, we calculate the atmospheric abundances of each element. To determine the mole fractions of the clouds, we found the abundance of potassium (1.23 $\times$ 10$^{-7}$ at solar metallicity, \cite{Burrows1999}), which is the limiting atom for KCl, at the specified atmospheric metallicity. We only include potassium chloride clouds, as ZnS (the other species that has a condensation curve within the temperature regime of the atmosphere of GJ 1214 b) is expected to be nucleation limited in its formation within the atmosphere of GJ 1214 b \citep{gao2018b}. However, it is possible that ZnS may heterogeneously nucleate on KCl cloud condensation nuclei (CCN) and produce a mixed cloud distribution \citep[e.g.,][]{gao2018b, Powell2024}. \cite{gao2018b} note that these mixed cloud distributions have less efficient condensational growth, resulting in smaller effective particle radii. Hazes may also serve as CCN, increasing cloud growth at the cost of decreasing haze particle density \cite{Yu2021}. However, detailed discussion of cloud---haze interactions is outside the scope of this work.

At present, there exists uncertainty in vertical cloud extent, as certain studies support the notion of compact clouds \citep{Ackerman2001}, while other dynamical models propose more pervasive vertical mixing \citep{Charnay2015a, Lines2018}. To parameterize this within our set of simulations, we choose three possible cloud set-ups (if temperatures remain cool enough for clouds to persist): cloud-free models, clouds with vertical extents set to approximately 8 scale heights (which we refer to as moderate clouds), and clouds that are allowed to form over the full 50 layers of the model, approximately 16 scale heights (which we refer to as extended clouds). Especially for higher metallicity atmospheres where the cloud condensation curves have been shifted to higher temperatures, our extended cloud parameterization may overestimate cloud coverage. The formation of deep atmosphere clouds and weak vertical mixing would result in the sequestration of clouds and cloud-free upper atmospheres. Assuming clouds can form at the bottom boundary layer, the moderate cloud cut-off corresponds to approximately 30 mbar, resulting in cloud-free upper atmospheres.

\subsection{Radiatively Active Photochemical Hazes}
In order to study how photochemical hazes influence the atmospheric dynamics of exoplanets, and how these changes manifest in observations, we add to the RM-GCM the capability of modeling photochemical hazes. Photochemical hazes are modeled as radiatively active aerosols that contribute an optical depth ($\tau$), single scattering albedo ($\varpi_0$), and asymmetry parameter ($g_0$) to each atmospheric layer.

\begin{figure*}[!htbp]
\begin{center}
\includegraphics[width=0.32\linewidth]{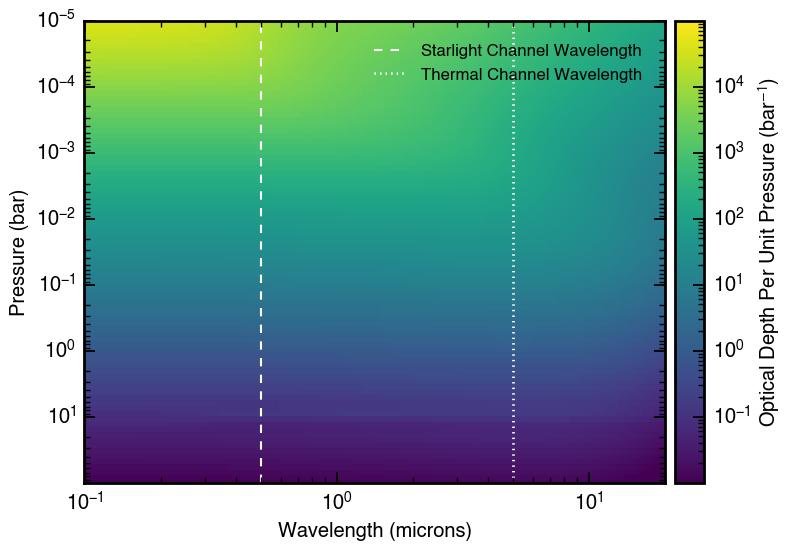}
\includegraphics[width=0.32\linewidth]{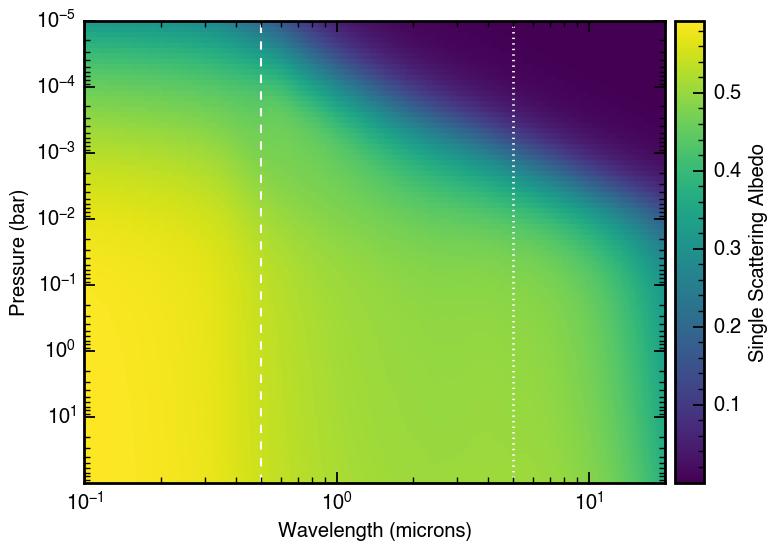}
\includegraphics[width=0.32\linewidth]{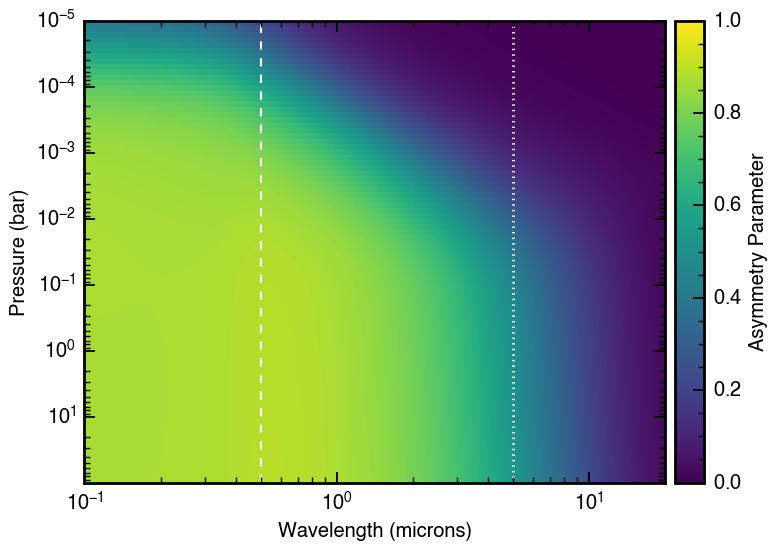}
\caption{The optical properties of the soot haze. From left to right, the panels show the optical depth per bar, the single scattering albedo ($\varpi_0$), and asymmetry parameter ($g_0$). The wavelengths that correspond to the starlight and thermal double gray channels are also marked in white. Importantly, the optical depth per bar shows how the soot haze can be a major contribution to layer optical depths in the upper atmosphere, especially at the shorter starlight wavelengths. For thermal wavelengths, the asymmetry parameter varies from close to 0 at low pressures near the top of the atmosphere (below approximately 0.1 mbar), and increasing to approximately 0.5  by a pressure of 100 mbar.}
\label{fig:soot}
\end{center}
\end{figure*}

We use the same haze properties as were calculated for the analysis by \cite{Kempton2023}, with the exception of the single scattering albedos for the extra-reflective hazes. These hazes were calculated using the Community Aerosol and Radiation Model for Atmosphere (CARMA) microphysical model \citep{Turco1979, Toon1988, Gao2018}. Refractive indices and Mie scattering properties (including scattering cross-sections, single scattering albedos, and scattering asymmetry parameters) were calculated as a function of height from the CARMA results. Haze cross sections were converted into haze optical depth per bar assuming a 1D temperature profile and hydrostatic equilibrium for an ideal gas. The radiative properties of the soot haze for a range of pressures and wavelengths are shown in Figure \ref{fig:soot}.

It is not straightforward how the haze production rate should change with metallicity, and theoretical and experimental results suggest that the relationship between metallicity and haze production rate is not monotonic. For example, \cite{lavvas2019} look at the mass flux of haze precursor molecules in photochemical models and find that the precursor mass flux drops to one-third of its initial value when metallicity increases from solar to 100x solar. However, increasing metallicity further, the haze production rate increases again for 1000x solar but decreases for 10000x solar. Different assumptions on which molecules contribute to haze production further increase the uncertainty. Experiments only cover the range $>$ 100x solar, but yield ambiguous results. In this work, we do not change haze density or optical properties as a function of metallicity.

Following \cite{Kempton2023}, we assume a column-integrated haze production rate of 10$^{-12}\ \text{g cm}^{-2}\ \text{s}^{-1}$ for soot, tholin, and extra-reflective soot hazes; however, we use different single scattering albedos for the extra-reflective soot hazes. In \cite{Kempton2023}, the extra-reflective soot hazes have single scattering albedos of 0.9999. Here, we instead double the single scattering albedos of the soot hazes. If this double produces a single scattering albedo greater than 0.9999, we truncate at 0.9999, as larger values can produce numerical instabilities within our radiative transfer scheme (and the single scattering albedo has a maximum physical value of 1).

For each layer, we calculated the weighted sum of the cloud, haze, and gas optical properties as

\begin{equation}
\kappa_{\rm total} = \kappa_{\rm gas} + \kappa_{\rm cloud} + \kappa_{\rm haze}, 
\end{equation}

\begin{equation} \label{eqn_w0}
\varpi_0 = \sum \frac{\tau_{\rm aerosol, i}}{\tau_{\rm tot}} \times \varpi_{0,i}, 
\end{equation}

\noindent and

\begin{equation} \label{eqn_g0}
g_0 = \sum \frac{\tau_{\rm aerosol,i}}{\tau_{\rm tot}} \times g_{0,i} .
\end{equation}

\noindent where the subscript $i$ refers to the species of cloud or haze. Hazes are ``painted on'', in that they do not form and dissipate at the model evolves. Instead, their presence is fixed by the precalculated optical depth and optical properties shown in Figure \ref{fig:soot}. We assume that hazes are globally uniform across every latitude and longitude. However, recent work by \cite{Steinrueck2021, Steinrueck2023} suggests different regimes of haze settling based on particle size.

From \cite{Steinrueck2021}, small particle hazes on HD 189733 b amass in mid-latitude vortices that extend from the nightside to the morning terminator. Conversely, large particle hazes settle quickly, resulting in more hazes on the evening terminator. The spatial distribution of hazes is critical to accurate atmospheric modeling, particularly on the dayside where upper atmosphere hazes can dominate whether incident starlight is scattered or absorbed. Recent observations have begun to probe morning-to-evening limb asymmetries in giant exoplanet atmospheres \citep[e.g.][]{ Murphy2024, Espinoza2024, Bell2024}. However, the complex interplay between haze formation, haze particle size distribution, and atmospheric dynamics means that there is still significant uncertainty in haze distribution.

Our uniform prescription has two benefits. First, it is computationally efficient, as modeling the formation of hazes from photochemical reactions in a 3D simulation is prohibitive. Secondly, the simplicity of the model makes it more easily explainable, allowing for a better understanding of the causal differences between models.

Within the RM-GCM, clouds and hazes are coupled to the radiative transfer scheme through their optical properties. Cloud and hazes both absorb and scatter light. In the double gray prescription, cloud and haze properties are evaluated at two distinct wavelengths: one for the starlight channel and one for the thermal channel. We opt to assess the radiative properties of the clouds and haze at 500 nm (for the starlight channel) and 5,000 nm (for the thermal channel), following \cite{Roman2019}. Furthermore, both cloud and haze radiative properties are relatively stable across small changes in wavelength (e.g., 500 nm vs 800 nm). Small changes in the wavelength at which we evaluated the double gray radiative transfer did not produce qualitative changes in model structure.

In order to compare different cloud and haze parameterizations of GJ 1214 b, we run a suite of models with varying haze properties, cloud extents, and metallicities. We simulate models with no hazes, soot hazes, tholin hazes, and extra reflective soot hazes. This final haze composition was simulated in order to compare to the extra reflective soot hazes preferred by the models that came closest to fitting the data in \cite{Kempton2023}. Overall we simulate all combinations of: three metallicities (1x, 30x, and 100x), three cloud extents (no clouds, moderate clouds, and extended clouds), and four different haze prescriptions (no haze, soot, tholin, and extra reflective soots), for a total of 36 simulations.

\subsection{Post-Processed Spectra and Phase Curves}
To compare the GCM outputs to observations, we post-process the simulations at a higher spectral resolution to calculate simulated emission spectra. We use the methodology described in previous work \citep{Zhang2017, Malsky2021}. We regrid the GCM models to constant altitude levels and interpolate pressure and temperature values to increase the number of vertical layers to 250. Next, we determine line-of-sight profiles and calculate the radiative transfer for each profile. For our radiative transfer scheme, we use the methodology outlined in \cite{Toon1989} to find both the thermal component and the reflected starlight component at each wavelength. We calculate atmospheric opacities using the \texttt{FastChem} equilibrium chemistry code \citep{stock2018fastchem, stock2022fastchem} with the following gas species: CO, H$_2$S, NH$_3$, CH$_4$, CO$_2$, H$_2$O. The use of these species was motivated by \cite{Moses2011} and \cite{millerricci2012}, as these species are predicted to be the most dominant opacity sources for approximately this pressure, temperature, and wavelength parameter space. We assume the atmosphere is dominated by molecular hydrogen and helium and use the set of collisionally induced opacities from \cite{Malsky2024b}. We post-process the GCM simulations at a resolution of $R=10^4$ for wavelengths between 5 $\mu$m and 12 $\mu$m. We also include the effects of radiatively active clouds and hazes following the methodology in \cite{Malsky2024b}. For the reflected starlight component, we assume the incident stellar flux takes the form of a blackbody with the stellar parameters listed in Table \ref{tab:planet_parameters}. To calculate simulated phase curves to compare to the data, which is measured as the planet-to-star flux ratio, we adopt the same \verb|PHOENIX| stellar spectrum as used in \cite{Kempton2023}, with a T$_{\mathrm{eff}}$ of 3250 K, log(g) = 5.0, and [M/H] = 0.2 \citep{Husser2013}. To create phase curves, we calculate emission spectra at 24 equally spaced phases.

\section{Results}\label{sec:results}
\begin{figure}[!htbp]
\begin{center}
\includegraphics[width=1.0\linewidth]{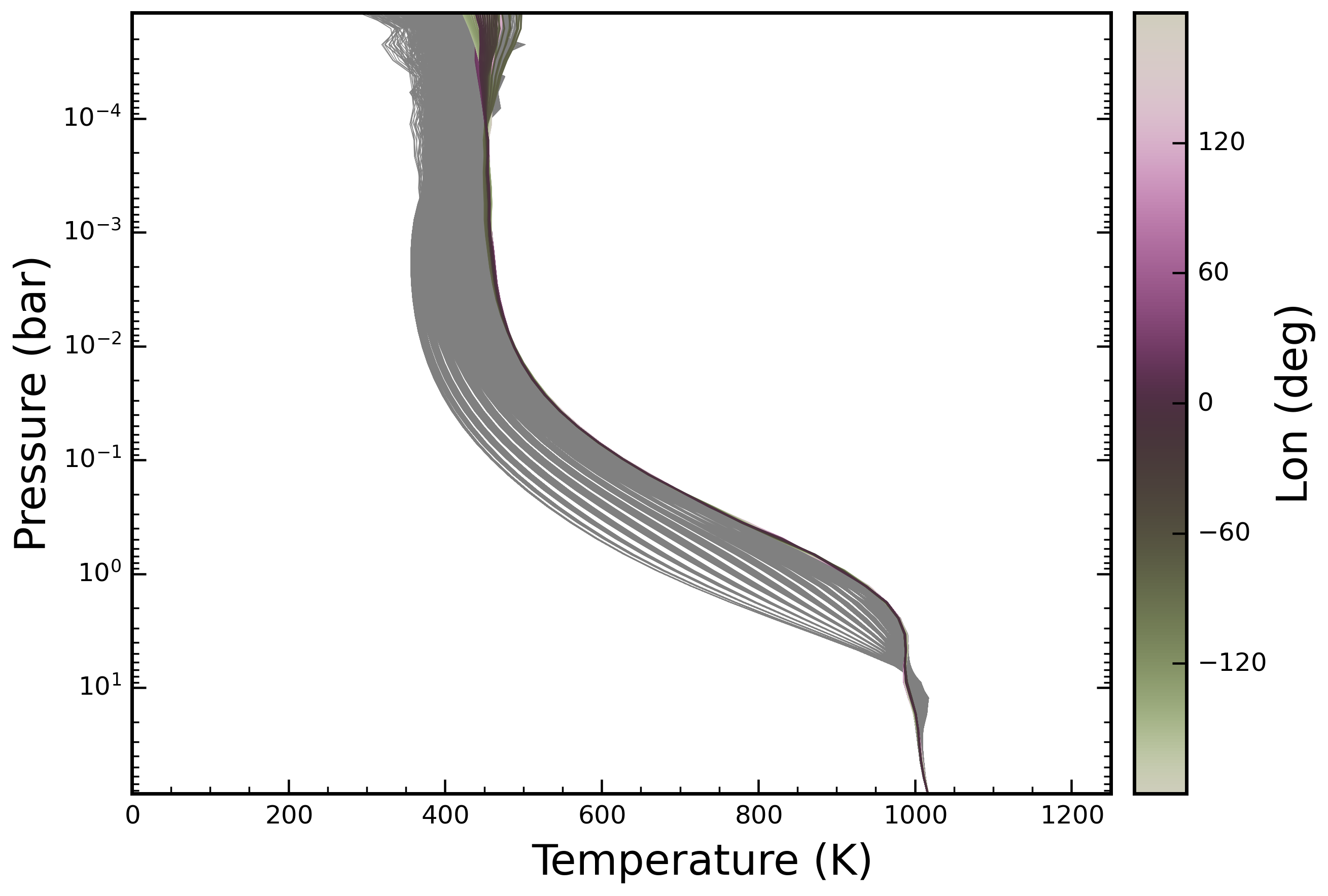}
\caption{Pressure-temperature profiles of a clear model of GJ 1214 b with no clouds or hazes and solar metallicity. Equatorial profiles are represented by colored lines, where the longitude is denoted by the color (zero corresponds to the substellar point). Pressure-temperature profiles for all non-equatorial profiles are represented by gray lines. Without aerosols, GJ 1214 b has equatorial day-night side temperature differences of less than 50 K.}
\label{fig:pt-clear}
\end{center}
\end{figure}

Without clouds or hazes, GJ 1214 b has upper atmosphere day-night temperature differences of less than 50 K and no thermal inversions across the 1x, 30x, and 100x solar models. Figure \ref{fig:pt-clear} shows the pressure-temperature profiles of the solar metallicity model with no clouds or hazes. These results are in general agreement with findings from previous works \citep[e.g.,][]{Menou2012, Kataria2014}. Namely, we find a super-rotating equatorial jet, small day-night temperature differences, similar atmospheric temperature structures, and deep atmosphere temperatures of approximately 1000 K.

\subsection{The Effect of Clouds or Hazes on Thermal Structure}
\begin{figure*}[!htbp]
\begin{center}
\includegraphics[width=0.99\linewidth]{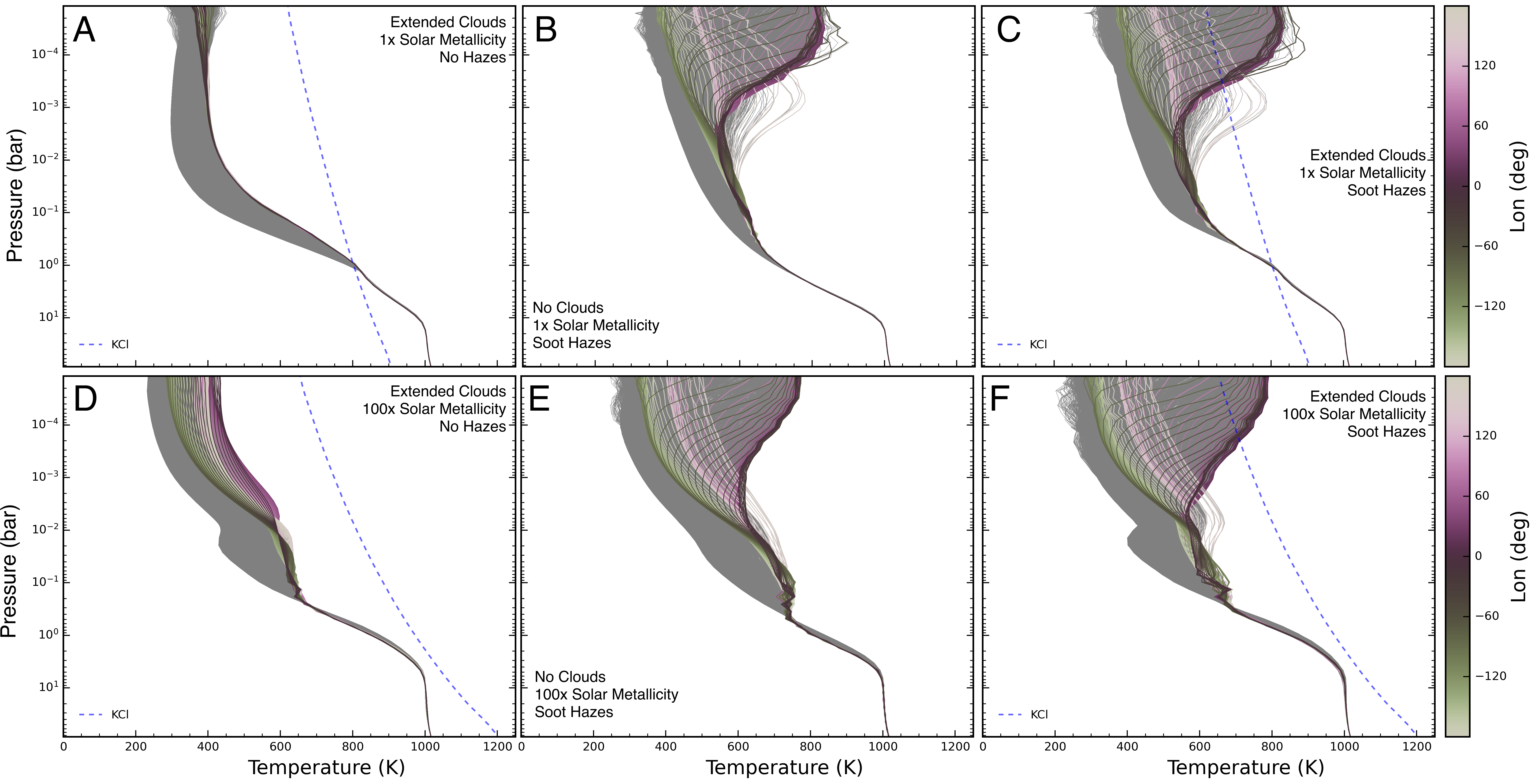}
\caption{Pressure-temperature profiles for models of GJ 1214 b with varying metallicity and aerosol distributions. From left to right, columns correspond to extended cloud models with no hazes, soot haze models with no clouds, and models with both extended clouds and soot hazes. The top row shows models at 1x solar metallicity, and the bottom row shows models at 100x solar metallicity. Equatorial profiles are represented by colored lines, where the longitude is denoted by the color (zero corresponds to the substellar point). Pressure-temperature profiles for all non-equatorial profiles are represented by gray lines. The condensation curves for the KCl clouds are shown as additional dashed blue lines. Atmospheric aerosols, particularly hazes, drive large day-night temperature differences.}
\label{fig:ptc}
\end{center}
\end{figure*}

The addition of aerosols --- either hazes or clouds --- increases day-night temperature differences, pole-equator temperature differences, and Bond albedos. KCl clouds and the hazes modeled here have dramatically different optical properties --- hazes are generally more absorptive, while clouds are reflective. While clouds (alone) result in day-night temperature differences of $\sim$100 K, soot hazes (alone) create temperature contrasts of more than 400 K (Figure \ref{fig:ptc}). Hazes result in larger day-night temperature contrasts than clouds, particularly at pressures less than 10$^{-3}$ bar. Additionally, only hazes result in thermal inversions of several hundred degrees.

Day-night and pole-equator temperature contrasts are larger in super-solar metallicity atmospheres (see Figure \ref{fig:ptc}). Increasing metallicity affects the atmospheric structure primarily through three changes: increased opacity, increased molecular weight, and decreased heat capacity \citep{zhang2017comp}. The effects of metallicity on the atmospheric circulation of GJ 1214 b has been detailed in other works (see \citealt{Kataria2014}, for example).

\begin{figure*}[!ht]
\begin{center}
\includegraphics[width=0.95\linewidth]{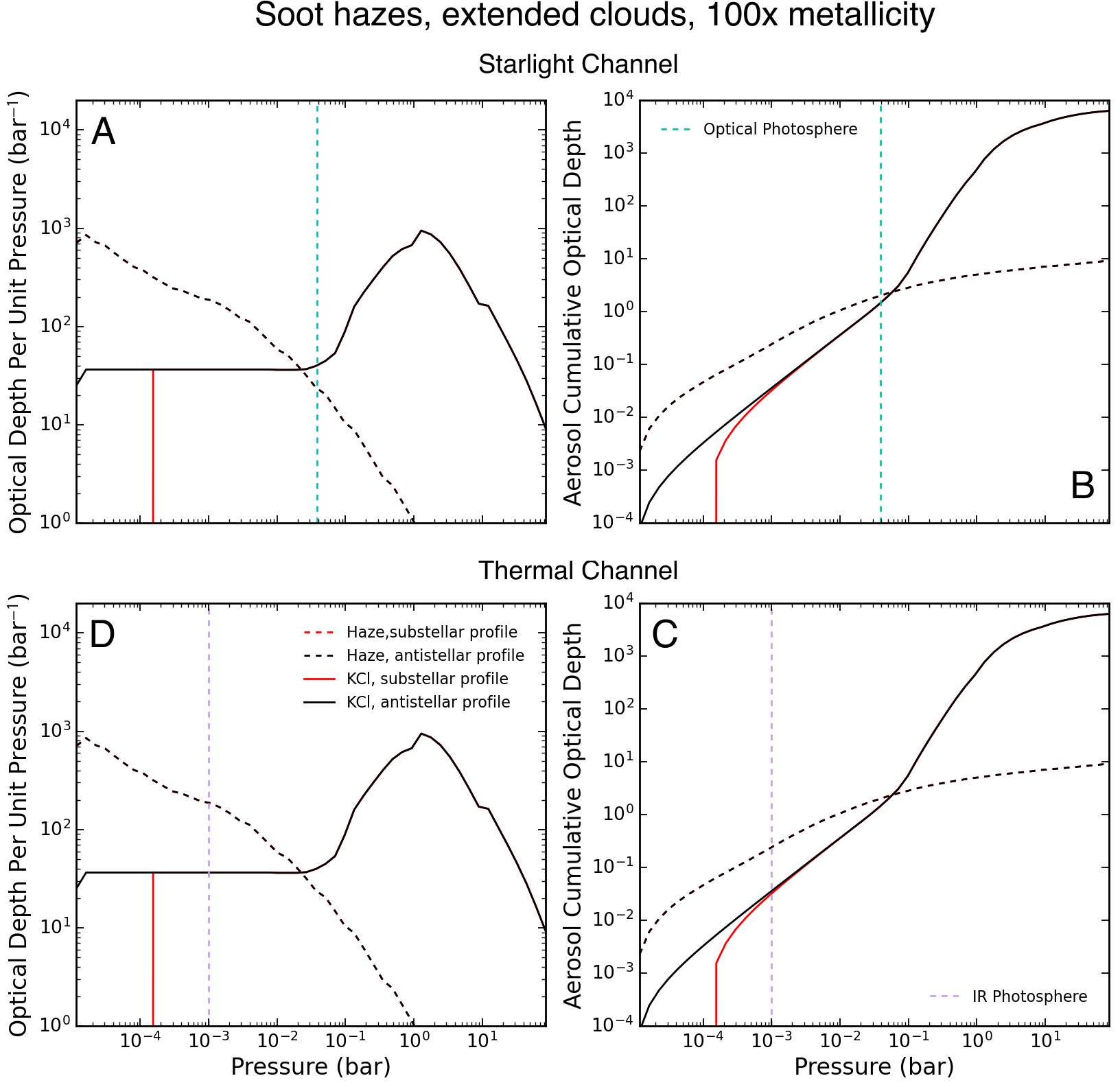}
\caption{Vertical aerosol profiles of GJ 1214 b simulated with extended clouds and soot hazes at 100x solar metallicity. Cloud and haze radiative properties are shown at 0.5 $\mu$m (top row) and at 5.00 $\mu$m (bottom row), corresponding to the starlight and thermal channels of the double gray radiative transfer. The red lines show the substellar profile, and the black lines show the antistellar profile. (For the hazes the substellar and antistellar profiles are identical.) Across all panels, the vertical lavender and turquoise lines mark the pressure level where the thermal and starlight photospheres would be, respectively, if no aerosols were present. The profiles reveal that hazes dominate the thermal channel at pressures lower than approximately 10$^{-1}$ bar, but that both hazes and clouds are optically thick at the optical photosphere.}
\label{fig:aerosol_profiles}
\end{center}
\end{figure*}

The day-night temperature differences caused by clouds increase as metallicity increases, due to the increase in available potassium atoms and the resulting increase in cloud density and opacity. In the upper atmosphere, clouds dominate over gas opacity for super-solar metallicities. Figure \ref{fig:aerosol_profiles} shows that for both the thermal and starlight channels, the integrated cloud optical depth reaches order unity at lower pressures than the gas contribution at 100x solar metallicity. The integrated gas optical depth in the thermal channel reaches unity at a pressure level of approximately 1 mbar for the 100x solar atmosphere based on our prescribed double gray absorption coefficients. In contrast, KCl clouds reach an integrated optical depth of order unity at 0.1 mbar for a 100x solar metallicity simulation with soot hazes and extended clouds. However, the dominance of clouds does not hold for the solar metallicity atmospheres, where cloud opacities are a factor of 100x less. This relation can also be seen in Figure \ref{fig:ptc}, where the effects of clouds on the thermal structure of the atmosphere only becomes significant at super-solar metallicities.

The balance between clouds, hazes, or gas dominating atmospheric opacity depends on pressure level, metallicity, temperature, assumed aerosol properties, and wavelength. For a 100x solar metallicity atmosphere at IR wavelengths, the opacity from hazes dominates over that of clouds at pressures less than $\sim$10$^{-2}$ (Figure \ref{fig:aerosol_profiles}). However, at visible wavelengths, clouds have larger opacities than hazes. The effect of aerosols on radiative transfer is further complicated by inhomogeneous cloud formation: the planet's dayside can reach temperatures hot enough to dissipate KCl clouds. The absorptive nature of soot hazes, combined with the thermal blanketing from clouds, results in the large day-night temperature differences and thermal inversions seen in Figure \ref{fig:ptc}. Because gas, hazes, and clouds all meaningfully contribute to atmospheric opacity in different pressure regions, the overall optical properties of the atmosphere are sensitive to changes from all three.

\begin{figure*}[!htbp]
\begin{center}
\includegraphics[width=1.0\linewidth]{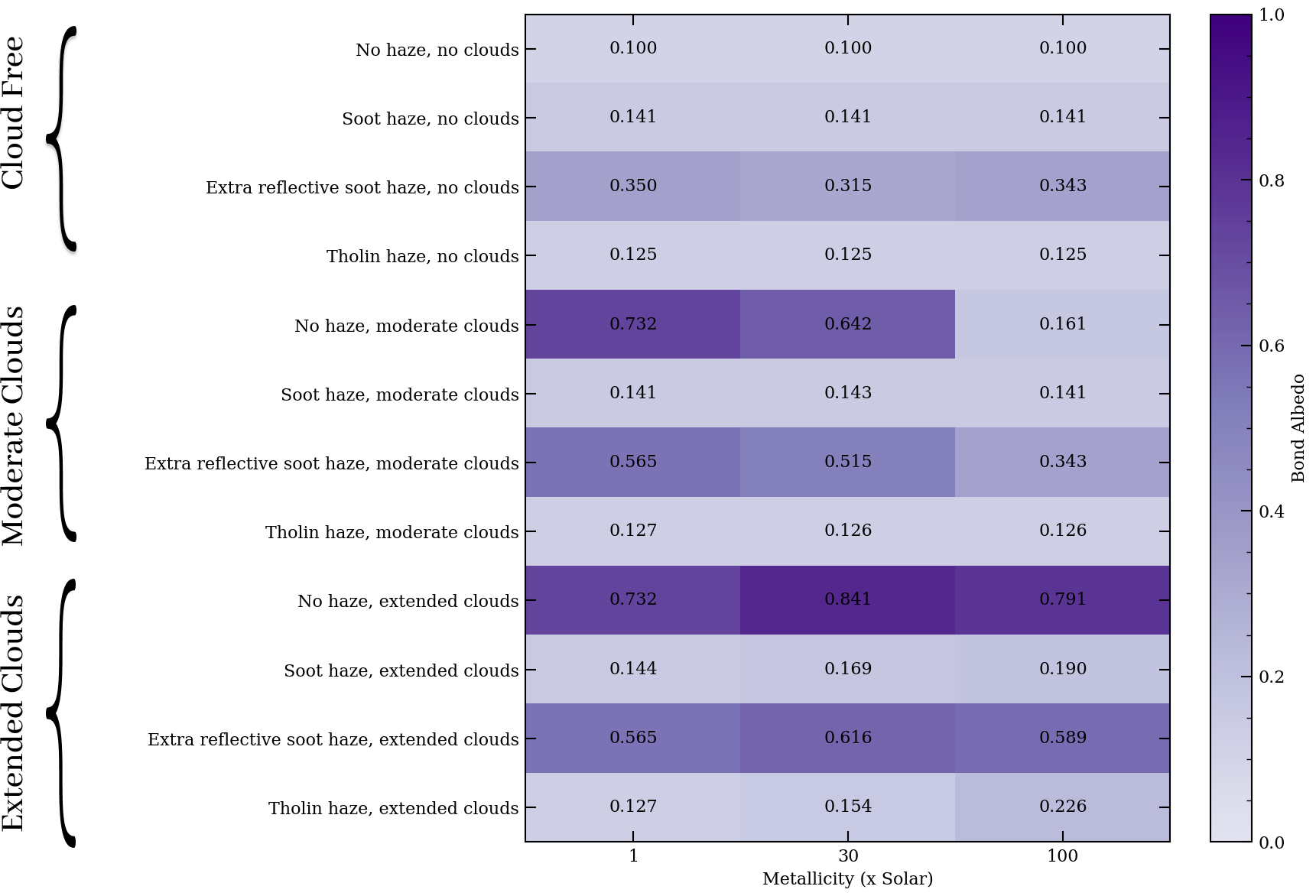}
\caption{The Bond albedos (the energy reflected by the planet divided by the total energy incident across all wavelengths) of GJ 1214 b models with aerosols for metallicities of 1x, 30x, and 100x solar. All models have an additional imposed top-of-the-atmosphere Bond albedo of 0.10, following \cite{Malsky2024}. KCl clouds or extra reflective soot haze models result in Bond albedos above 0.3.}
\label{fig:bond_albedos}
\end{center}
\end{figure*}

Radiatively active clouds have a major impact on the starlight reflected by GJ 1214 b's atmosphere, increasing the Bond albedo to 0.73, 0.84, and 0.79 for the 1x, 30x, and 100x solar metallicity cases, respectively, from the nominal prescribed clear-atmosphere Bond albedo of 0.1 (taken as an approximate and representative global Bond albedo to represent the effect of Rayleigh scattering in a clear atmosphere; see \citealt{Malsky2024}). The Bond albedos of all our models are shown in Figure \ref{fig:bond_albedos}. Thick KCl clouds in the planet's upper atmosphere result in extremely reflective atmospheres. This leads to cooler upper atmospheres for cloudy models compared to clear models ($\sim$300 K vs $\sim$450 K), as shown by comparison between Figures \ref{fig:pt-clear} and \ref{fig:ptc}.

Because of their large single-scattering albedos, clouds result in more reflected starlight than hazes. Figure \ref{fig:bond_albedos} shows the resulting Bond albedos of our set of models. Soot haze models result in Bond albedos of approximately 0.14, compared to Bond albedos above 0.5 for some haze-free KCl cloud models. The Bond albedo of our soot haze models did not change more than a few percent from 1x to 100x solar metallicity. This is because the soot haze opacity dominates over gas opacity in the upper atmosphere, resulting in nearly constant global single-scattering albedos for similar models with different metallicities. However, it is worth noting that this result is tied to our assumption of the same amount of haze production for every atmospheric metallicity. Because single scattering albedo is calculated as a weighted average of the gas and aerosol contributions, at a certain point, clouds saturate the single scattering albedo, and further increases in abundance do not further enhance atmospheric reflectivity.

\begin{figure*}
\begin{center}
\includegraphics[width=0.9\linewidth]{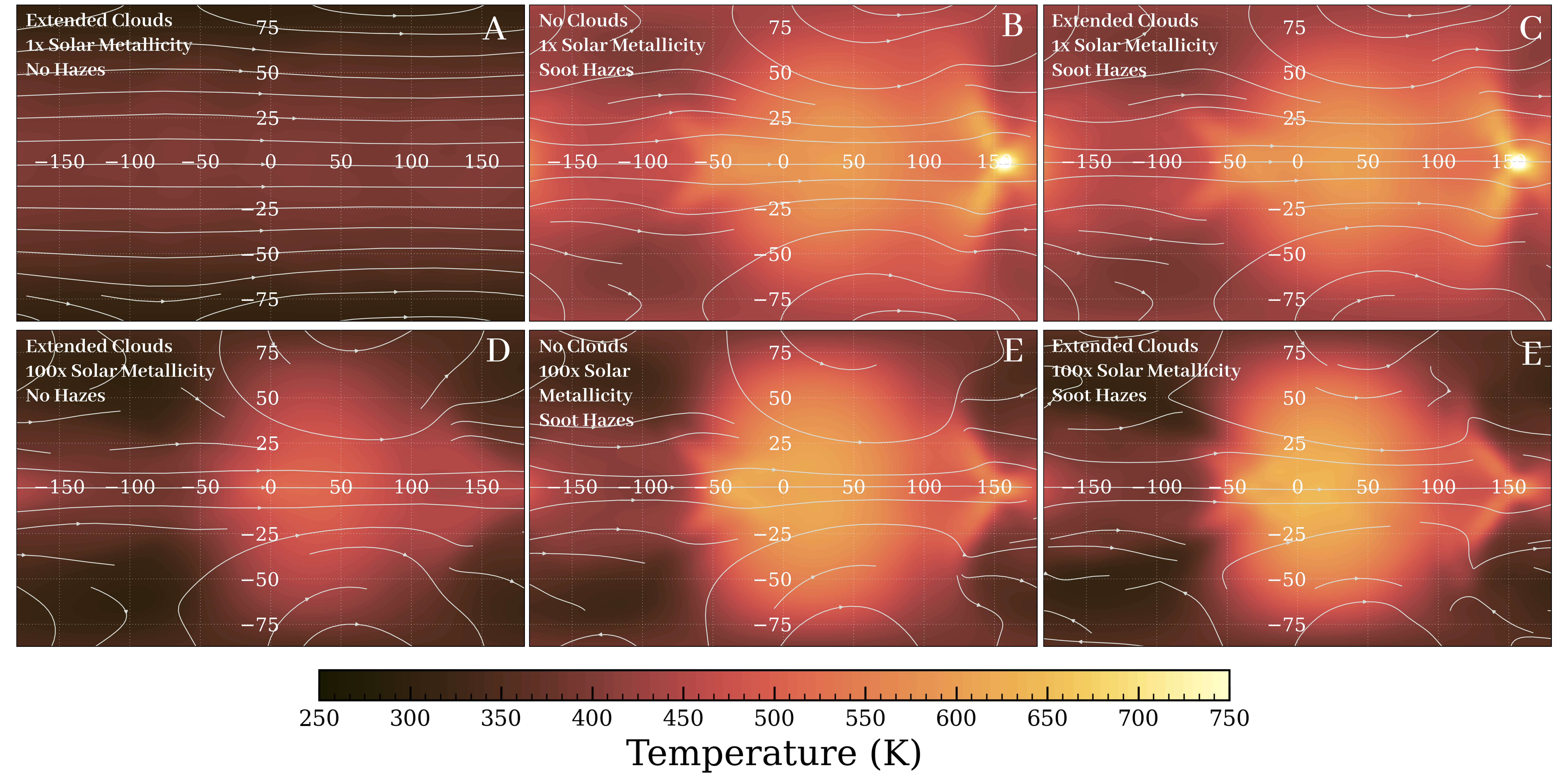}
\caption{The temperature maps and wind structure (white arrows) of different GJ 1214 b models around a pressure of 1 mbar. The panels show 1x solar metallicity (top row) and 100x solar metallicity (bottom row). From left to right, the columns show extended cloud models, models with soot hazes, and models with both extended clouds and soot hazes. Hazes absorb more starlight by this pressure level, resulting in larger day-night temperature differences.}
\label{fig:isobar}
\end{center}
\end{figure*}

Figure \ref{fig:isobar} shows temperature and wind maps of GJ 1214 b at 1 mbar, for various aerosol compositions. Soot hazes increase day-night temperature differences, equatorial wind speeds, and equator-to-pole temperature differences. At this pressure level, cloudy models exhibit a relatively homogeneous longitudinal temperature structure of approximately 375 K, with 100 K equator-to-pole temperature differences. Due to thicker cloud coverage, the 100x solar metallicity cloud model has a hotter dayside, a colder nightside, and approximately 200 K equator-to-pole temperature differences. In contrast, the soot haze model shows large equator-to-pole and day-night temperature differences for both the 1x and 100x metallicity simulations. At 1 mbar, the tholin and soot haze models result in the fastest eastward winds, generally between 2 and 3 km s$^{-1}$. The clear and haze free models have slower winds, with eastward wind speeds between 1 and 2 km s$^{-1}$. The absorptive hazes create day-night temperature contrasts, which then drive the global upper atmosphere circulation.

The haze models also exhibit dynamics-driven, chevron-shaped temperature features \citep[e.g.,][]{Showman2009, Rauscher2010, Tsai2014, Showman2011, Beltz2022}. Additionally, the hotter temperature regions are shifted eastward from the dayside irradiation due to advection with the eastward circulation. When clouds and hazes are present together, the isobaric structure resembles that of the haze-only models.

Similar to previous works \citep{Menou2012, Kataria2014, Charnay2015a}, the clear solar-metallicity simulations result in two mid-latitude jets with peak speeds of $\sim$1.5 km s$^{-1}$. As metallicity increases, molecular weight, molar heat capacity, and day-night temperature contrasts increase. Additionally, the higher metallicity models have increased opacities, resulting in shallower heating, a decrease in the strength of the high-latitude jets, and an increase in the speed of the equatorial jet \citep{Kataria2014}. For the 100x solar metallicity models, the circulation transitions to a single equatorial jet with a peak speed of $\sim$2.0 km s$^{-1}$, qualitatively consistent with previous models. However, we do not find a narrowing of the jets with increasing metallicity associated with a decreasing Rossby deformation radius \citep{zhang2017comp}. Hazes (and, to a lesser extent, clouds) increase day-night temperature differences and contribute to the formation of a broad equatorial jet. Further analysis of the effects of photochemical hazes on the circulation and dynamics of GJ 1214 b is presented in \SteinrueckEtAlT.

{\subsection{Comparison to the SPARC/MITgcm}\label{sec:SPARC_compare}
\cite{Kempton2023} included analysis from a set of GCMs simulated using the SPARC/MITgcm. A subset of the simulated observables from these models is included in subsequent sections of this paper. Here, we detail several important underlying differences between the two GCMs. Briefly, the SPARC/MITgcm solves the primitive equations on a cubed-sphere grid with correlated-k distribution radiative transfer \citep{Adcroft2004, Showman2009, Kataria2014}. Differences arising from the radiative transfer of the SPARC/MITgcm have a large (but complex) impact on the atmospheric heating/cooling, and the pressure level of the wavelength-dependent photosphere. Other differences between the two GCMs (including drag prescription and stellar model) are outlined in \cite{Kempton2023}.

Qualitatively, there is agreement between the solar metallicity-clear models simulated with the RM-GCM and the SPARC/MITgcm. At 1 mbar, the SPARC/MITgcm (RM-GCM) models have $\sim$50 K ($\leq$25 K) day-night temperature differences and $\sim$150 K ($\sim$100 K) equator-to-pole temperature differences. Both models show similar overall upper atmosphere temperatures. Similar aerosol-free models of GJ 1214 b have been extensively covered in literature, so the details here are only discussed in context of interpreting the most recent phase curve.

There are larger differences between the super-solar metallicity RM-GCM and SPARC/MITgcm models compared to the solar models. For 100x solar metallicity, the SPARC/MITgcm results in $\sim$250 K day-night and $\sim$150 K equator-to-pole temperature differences, whereas the RM-GCM models produce $\sim$100 K day-night temperature differences and $\sim$200 K equator-to-pole temperature differences. The 100x clear SPARC/MITgcm profiles also generally decrease in temperature at pressures less 10$^{-3}$ bar, whereas the RM-GCM models are isothermal. We attribute these differences largely to the multiwavelength opacity calculations in the SPARC/MITgcm.

\subsection{Comparison to JWST Spectral Phase Curve}\label{sec:JWST_compare}
We compare the results of our GCMs to the original JWST data and our reanalysis (see Appendix). Figure \ref{fig:phasecurves} shows the observed white light phase curves from \cite{Kempton2023} and simulated GCM phase curves. Comparing the reanalysis to the \cite{Kempton2023} analysis, we find an increase in the nightside flux (the most uncertain part of the original analysis), as well as a decrease in the Bond albedo. We compare our GCMs to these analyses, as well as the multi-wavelength, hazy GCMs presented with the original analysis.

Following the methodology from \cite{Kempton2023}, we calculate the Bond albedo from the spectroscopic phase curve. Using the full dataset from 5-12 {\micron} (with 0.5 {\micron} bins) we find a Bond albedo of 0.42 $\pm$ 0.11. Comparatively, \cite{Kempton2023} found a Bond albedo of 0.51 $\pm$ 0.06 from the 5-12 {\micron} spectroscopic phase curve. We also calculate the Bond albedo using only the channels from 5.0 - 10.5 {\micron}, as we find that these data are more reliable (see Appendix). We assume that 54\% $\pm$ 4\% of the planet luminosity is with the 5-12 {\micron} band (identically to \citealt{Kempton2023}), and that 49\% $\pm$ 4\% of the planet luminosity is with the 5-10.5 {\micron} band. Using this wavelength range with 0.25 {\micron} bins we find a Bond albedo of 0.46 $\pm$ 0.13. These new estimates are slightly lower but consistent with the original value, within the errors.

Aerosol-free models result in flat phase curves that overestimate total planetary flux (Figure \ref{fig:phasecurves}). These behaviors are due to the fact that these models have small underlying day-night temperature contrasts, and low Bond albedos. The RM-GCM and SPARC/MITgcm models both result in relatively similar phase curves for the 1x solar clear simulations, though the SPARC/MITgcm model exhibits a somewhat larger phase curve amplitude than the RM-GCM model. This difference is amplified for the 100x solar metallicity model. However, the trend for both GCMs is similar: an increase in phase curve amplitude with increased metallicity due primarily to changes in atmospheric opacity, mean molecular weight, and heat capacity.

Figure \ref{fig:phasecurves} shows that the 100x solar clear model has a greater integrated phase curve flux than the 1x solar clear model. As required by energy conservation, the total thermal emission integrated over all phases should be identical between clear models, and this is the case for the broadband thermal emission within the GCM. However, the photosphere levels differ between the double gray GCMs and the post-processing with opacity sampling, as the opacities are not fully self-consistent. For the 100x solar model, the IR photosphere is deeper ($\sim$1-20 mbar, in line with 1-D radiative equilibrium models) in the post-processing compared to the double gray GCM ($\sim$ 1 mbar). At this deeper pressure, the hotter temperatures result in greater flux. The importance of the gas-only photosphere is lessened for the aerosol models, where clouds or hazes become optically thick over a wide wavelength range at pressures of $\sim$1 mbar. However, this is not the case for the intermediate cloud models, where clouds do not form in the upper atmosphere, and the IR photosphere pressure level is set by the gas opacity. The photosphere pressure level is inherently wavelength-dependent and cannot be perfectly captured by the double gray prescription. Additionally, Figure \ref{fig:phasecurves} shows only the 5-12 $\mu$m band, and a portion of the emission is outside this band (especially at shorter wavelengths, as shown in Figure \ref{fig:spectra}), and varies between the 1x, 30x, and 100x metallicity models.

Aerosols increase phase curve amplitudes and increase planetary Bond albedos --- a trend necessary to explain the observed phase curve data. We find this trend for both the RM-GCM and the SPARC/MITgcm models (see Figure \ref{fig:phasecurves}). We do not expect the RM-GCM and SPARC/MITgcm models to produce identical results due to differences in model architecture --- particularly the multiwavelength radiative transfer scheme within SPARC/MITgcm, compared to the double gray version we use in the RM-GCM models presented here.

No GCM within our set matches the phase curve amplitude, offset, and peak from the observation. However, clouds, hazes, and super-solar metallicities help explain the observation and aid in characterizing the underlying atmosphere of GJ 1214 b. Although clouds and hazes both increase the Bond albedo, clouds are particularly reflective. Although hazes can dominate over clouds in the upper atmosphere, both clouds and hazes affect the observables in concert. Figure \ref{fig:phasecurves} shows that the addition of extended clouds to the solar metallicity, extra reflective soot models decreases the nightside flux by $\sim$10\%. In agreement with previous studies, we find that clouds increase phase curve amplitudes and decrease phase curve offsets, while hazes can particularly strongly influence phase curve amplitudes.

Figure \ref{fig:spectra} shows the post-processed GCM simulations compared to the observed day- and night-side emission spectra data from \cite{Kempton2023}. Similar to the analysis in \cite{Kempton2023}, we find no GCM simulation that agrees with the observational data within uncertainties. On the dayside, models with soot or extra reflective soot aerosols are reasonably good fits. The clear solar metallicity model over-predicts the nightside flux, as well as dayside fluxes from $\sim$8.5 $\mu$m to $\sim$10.25 $\mu$m. Only the extended cloudy models have low enough nightside temperatures to account for the low nightside emission. Models with aerosols come the closest to suppressing the nightside emission enough to match the data. Since none of the models match the data well, instead we can try to focus on overall trends in how the metallicity, aerosol type, and aerosol distribution influence the planet's emission.

Extended cloud models show the lowest planet-to-star flux ratios. This is in line with these models' extremely large Bond albedos (in excess of 0.7) --- the highest of all the simulations presented here. The extended cloud models produce similar emission spectra for both the day and night sides, with only slightly larger dayside fluxes. The extended cloud models produce emission spectra that are qualitatively a good fit to the nightside spectra, but dramatically under-predict dayside flux. Hazes result in large day-night flux differences, but generally over-predict white light fluxes. Additionally, both hazes and clouds lead to muted spectral features compared to the clear atmosphere models. Our solar metallicity model---with hazes whose single scattering albedos are enhanced by a factor of 2 and also extended clouds---produces the closest match to the data from \cite{Kempton2023}. However, even this model over-predicts flux ratios at wavelengths less than approximately 8 $\mu$m for both the day and night side.

\begin{figure*}
\begin{center}
\includegraphics[width=0.95\linewidth]{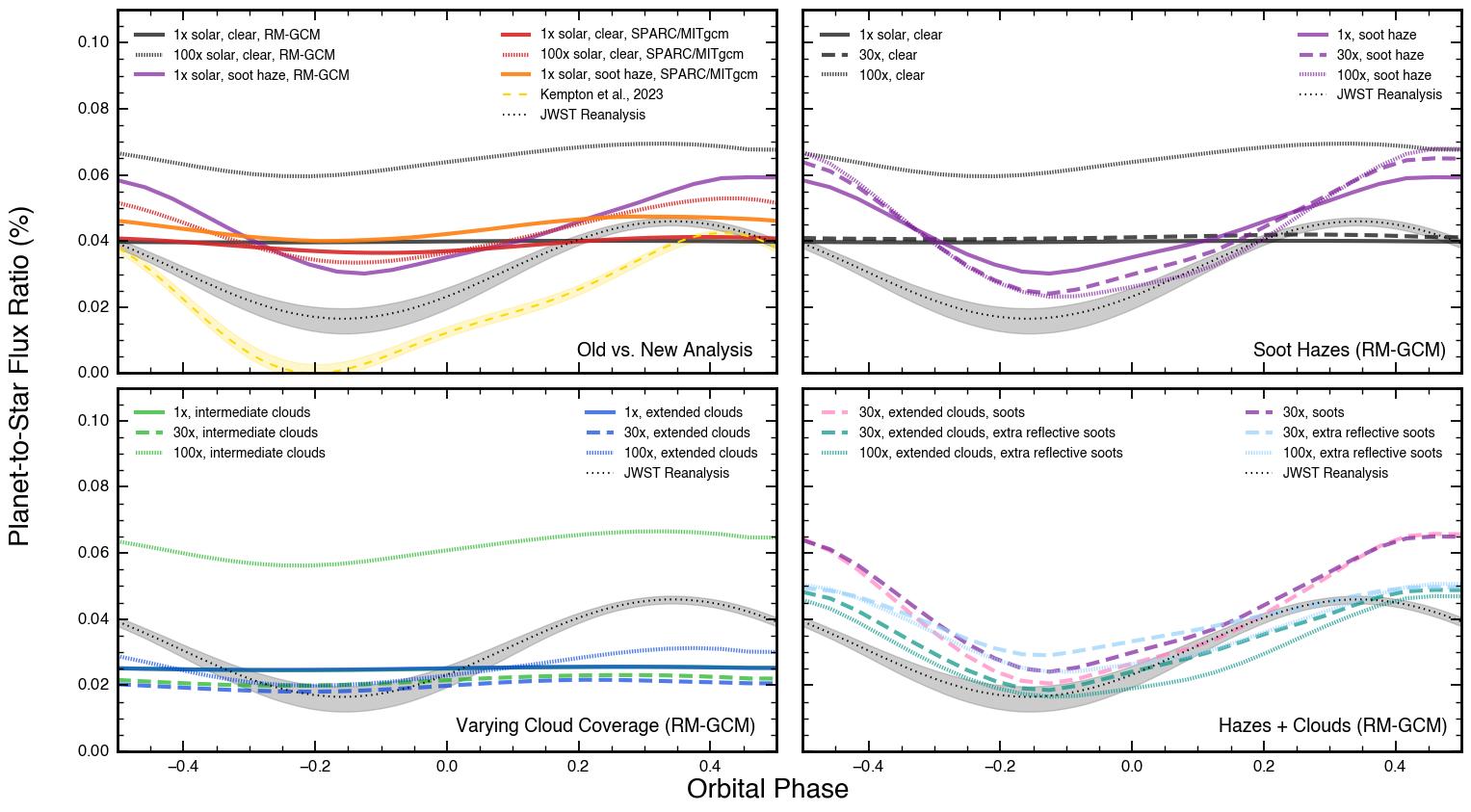}
\caption{Phase curves of GJ 1214 b, integrated from 5-12 $\mu$m. The dashed gray line and surrounding shaded region show a fit (with an imposed positive flux requirement) to the observed data from \cite{Kempton2023} and the corresponding 3$\sigma$ uncertainty. The black dotted line and surrounding shaded region show a reanalysis of the same data, with the methodology presented in the Appendix of this work. In the top left panel, dashed lines show clear and hazy SPARC/MITgcm models from \cite{Kempton2023}. In all other panels, sets of RM-GCM models were chosen to show trends with metallicity, cloud coverage, and hazes. In agreement with the original analysis, we find that the presence of atmospheric aerosols (and likely clouds and hazes together) are required to explain the large observed phase curve amplitude and lower overall planetary flux, but no model is currently able to match the data.}
\label{fig:phasecurves}
\end{center}
\end{figure*}

\begin{figure*}
\begin{center}
\includegraphics[width=0.95\linewidth]{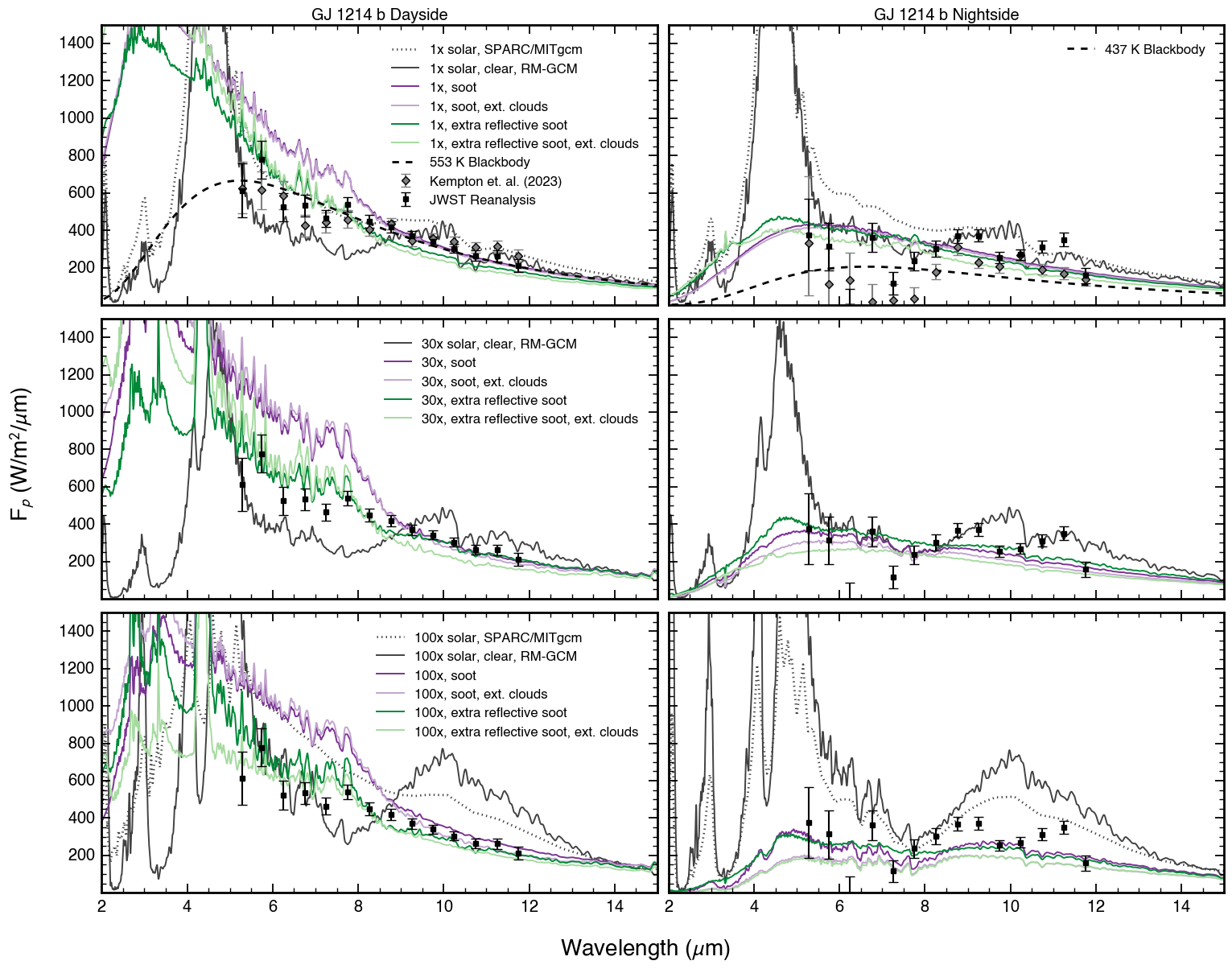}
\caption{Emission spectra from the dayside (left) and nightside (right) of GJ 1214 b. The gray diamonds show the observed spectra from \cite{Kempton2023}. The black squares show the reanalysis. The black dashed lines show a 553 K blackbody (right) and a 437 K blackbody (left), the best fit values from \cite{Kempton2023}. We use the same  PHOENIX \citep{Husser2013} stellar model as in \cite{Kempton2023} to convert the SPARC/MITgcm models to planetary flux. Simulated spectra are also shown for 1x solar metallicity (top row), 30x (middle row), and 100x (bottom row). The dotted lines show SPARC/MITgcm models, while the solid lines show RM-GCM models for varying aerosol parameterizations (convolved down to R = 100). An atmosphere with a combination of clouds, hazes, and a higher metallicity atmosphere best fits the day and nightside emission spectra.}
\label{fig:spectra}
\end{center}
\end{figure*}

\section{Conclusions}\label{sec:conclusions}
In this work, we present novel 3D models of clouds and hazes in the atmosphere of GJ 1214 b and compare a suite of models to a reanalysis of the MIRI phase curve. The addition of hazes to the RM-GCM enables more diverse modeling of cooler exoplanets that may have both clouds and hazes. A number of GCMs include clouds \citep[e.g.,][]{Lee2016, Parmentier2016, Helling2016, Amundsen2016, Lines2018, Roman2019, Steinrueck2021, Parmentier2021, Christie2021, Komacek2022} and fewer include hazes \citep{Steinrueck2021, Steinrueck2023}. This work marks the first time that both clouds and hazes are included in a published exoplanet GCM.

In agreement with previous hazy GCMs presented by \cite{Kempton2023} using the SPARC/MITgcm, we find that models with higher metallicities and/or with hazes produce larger day-night temperature differences. The presence of clouds in our models also increases day-night temperature differences, but to a lesser extent than hazes. Clouds alone do not produce the large temperature inversions found in the hazy models \citep[see also][]{Steinrueck2023}, due to the assumed aerosol properties, which have the clouds being more reflective and the hazes being more absorptive.

This work also presents a reanalysis of the \cite{Kempton2023} GJ 1214 b phase curve data, and identifies a wavelength-dependent systematic in the flux time series that, when corrected, results in a brighter nightside and a lower estimate for the planet's Bond albedo. The overall interpretation of the phase curve remains consistent with the results from \cite{Kempton2023}: models must have high metallicity and substantial aerosols in order to agree with observations. We find that the models that are closest to the data are those with both clouds and hazes. Due to the strongly reflective nature of the clouds combined with the absorptive properties of the hazes, these atmospheres show both the large albedo and day-night flux differences indicated by the observations. Trends within our models show that both clouds and hazes may be necessary to match the thermal structure predicted by \cite{Kempton2023}.

None of the models in this work, nor the ones from \cite{Kempton2023}, can match the amplitude and shift of the broadband phase curve, but there are now more models that better match the overall global emission from the planet. Hazes alone are not reflective enough to match the phase curve of GJ 1214 b. Here, we only test a haze production rate of 10$^{-12}\ \text{g cm}^{-2}\ \text{s}^{-1}$, but higher production rates in the atmosphere of GJ 1214 b may result in better fits \citep{Gao2023}. The largest haze-only Bond albedo from our models was that of the extra-reflective soot haze simulation, with a value of 0.34. However, clouds $\textit{and}$ hazes can produce the amount of reflected starlight required. The extra-reflective soot hazes with moderate or extended clouds result in Bond albedos of between 0.32 and 0.84. However, the soot and tholin haze models are too absorptive to match the measured Bond albedo. Clouds alone do produce sufficiently high Bond albedos (0.16 - 0.84), but are not able to reproduce the large flux contrast seen in the phase curve and the expected temperature structure of GJ 1214 b.

Aerosols are complex and expensive to model --- particularly in three-dimensional simulations. Uncertainties in cloud density, vertical mixing, haze optical properties, haze production rate, and other physical processes mean that other scenarios could be consistent with the available data. The mysteries of GJ 1214 b are slowly being revealed, but more data and more models will be needed to more conclusively determine the composition of its atmosphere.

\section*{Acknowledgments}
IM would like to thank MC for providing editorial suggestions on drafts of this manuscript. This research was supported in part through computational resources and services provided by Advanced Research Computing at the University of Michigan, Ann Arbor. This work received financial support from the NASA Exoplanets Research Program Grant \#80NSSC22K0313 and the Heising-Simons Foundation. M.S. and M.Z. acknowledge support from the 51 Pegasi b Fellowship, funded by the Heising-Simons Foundation. This work is based on observations made with the NASA/ESA/CSA James Webb Space Telescope. The data were obtained from the Mikulski Archive for Space Telescopes at the Space Telescope Science Institute, which is operated by the Association of Universities for Research in Astronomy, Inc., under NASA contract NAS 5-03127 for JWST. These observations are associated with program \#1803 (catalog DOI: 10.17909/qe3z-qj40, https://doi.org/10.17909/qe3z-qj40). Support for Program number 1803 was provided through a grant from the STScI under NASA contract NAS5-03127. This research was carried out at the Jet Propulsion Laboratory, California Institute of Technology, under a contract with the National Aeronautics and Space Administration (80NM0018D0004).

\begin{appendix}
\section{A Re-analysis of the MIRI Phase Curve}
In the nearly two years since the start of science operations, the transiting exoplanet community has learned many lessons on how to properly reduce JWST time-series data.  As a result, we reanalysed the GJ~1214b MIRI phase curve data from GO-1803 (co-PIs Bean \& Kempton).  While we applied many of the same techniques used in the original reduction of GJ~1214b, we uncovered a previously-unidentified, possible systematic that, if real, drastically alters the measured nightside emission.  This, in turn, impacts the inferred heat redistribution efficiency and albedo.  As noted by \cite{Kempton2023}, the nightside constraint was the least robust aspect of the original analysis; this remains true for our re-analysis.

\subsection{Data Reduction}
We used the \eureka~pipeline for data reduction and light curve fitting \citep{Bell2022}.  Starting with the \textit{\_uncal.fits} files, we processed the data through Stages 1 and 2 using mostly the default settings.  Importantly, we set the jump rejection threshold to 7$\sigma$.
For Stage 3, we processed all 50 segments in a single batch, thus ensuring a common median frame across all segments and no unwanted flux offsets between segments.  We performed a double-iteration, $4\sigma$ outlier rejection along the time axis for each pixel in the background region.  We then applied column-by-column background subtraction with an inner edge of 24 pixels from the trace.  During this step, we applied a $2.5\sigma$ outlier threshold to flag a few obvious hot pixels.  We adopted an aperture half-width of 3 pixels (7 pixels total width) and used the aforementioned median frame as weighting for optimal spectral extraction.

In Stage 4, we produced two different sets of spectroscopic light curves.  The first consisted of 22 spectroscopic channels from 5.0 -- 10.5 {\microns} (0.25{\micron} bins) and the seconded consisted of 14 spectroscopic channels from 5.0 -- 12.0 {\microns} (0.50{\micron} bins).  The former uses a more reliable wavelength range; however, the latter matches the spectroscopic channels used in our original analysis \citep{Kempton2023}.  We mask two pixel rows (156 and 352) and apply a $4\sigma$ outlier rejection to binned light curves using a 30-pixel rolling median.

\subsection{Light-Curve Fitting}
We used ExoTiC-LD \citep{Grant2022} and the Stagger grid \citep{magic2015stagger} to compute quadratic limb-darkening parameters (T$_{\mathrm {eff}}$=3250~K, log(g)=5.03, Z/H=0.29) for our light-curve fits.
We manually clipped the first 500 integrations from the start of the visit where MIRI's exponential ramp is steepest.  We also clipped the first $\sim75$ integrations from the start of each telescope re-pointing (integrations 4248--4322, 8567--8642, 12887--12962, 17207--17285, and 21527--21600) due to elevate flux levels.  Finally, we clipped integrations 9880--10550 (MJD$_{TDB} \sim 1.45$) to remove a possible flaring event just prior to transit.

We fit the transit and both eclipses using \texttt{batman} \citep{Kreidberg2015b} and fit the phase curve using a double-sinusoidal function of the form:
\[
PC = 1 + A_1 (\cos(\phi)-1) + A_2 \sin(\phi) +
         A_3 (\cos(2\phi)-1) + A_4 \sin(2\phi)
\]
where $A_N$ is the amplitude of the $N^{th}$ component and $\phi$ is the orbital phase relative to secondary eclipse.  We apply a linear trend in time to account for any stellar variability over the course of the 1.7-day observation and a falling exponential function in time to fit the detector systematics.  Since the steepness of the falling exponential is poorly constrained in the spectroscopic light curves, we apply a Gaussian prior using the median and $1\sigma$ uncertainties from the white light curve fit.

Unique to our re-analysis, we note a sudden, wavelength-dependent offset in the measured flux at MJD$_{TDB}$ = 1.0278 days.  We fit this systematic with a step function.  Figure \ref{fig:stepsize} plots the measured step size as a function of wavelength for three different analyses.  In each case, the step size is consistent with zero at the shortest MIRI wavelengths and increases exponentially at longer wavelengths.  The timing of the step function is unfortunate, in that it occurs near the minimum in planet flux and where there are significant systematics (see Figure \ref{fig:StepNoStep}).  Adding the step function results in a smaller phase curve amplitude and warmer nightside.

\begin{figure}
    \centering
    \includegraphics[width=0.6\linewidth]{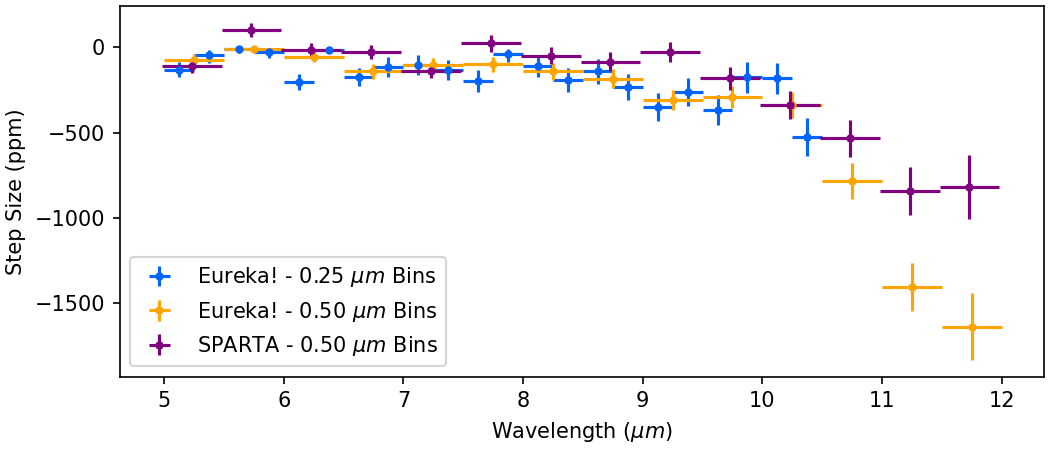}
    \caption{Measured step size (i.e., flux offset) versus wavelength for three different re-analyses.  All three re-analyses detect a jump in flux to high confidence at longer wavelengths; however, they disagree on the magnitude of the step size at $>11$ {\microns} where the measured flux and resulting phase curve fits are least reliable.}
    \label{fig:stepsize}
\end{figure}

\begin{figure}
    \centering
    \includegraphics[width=0.6\linewidth]{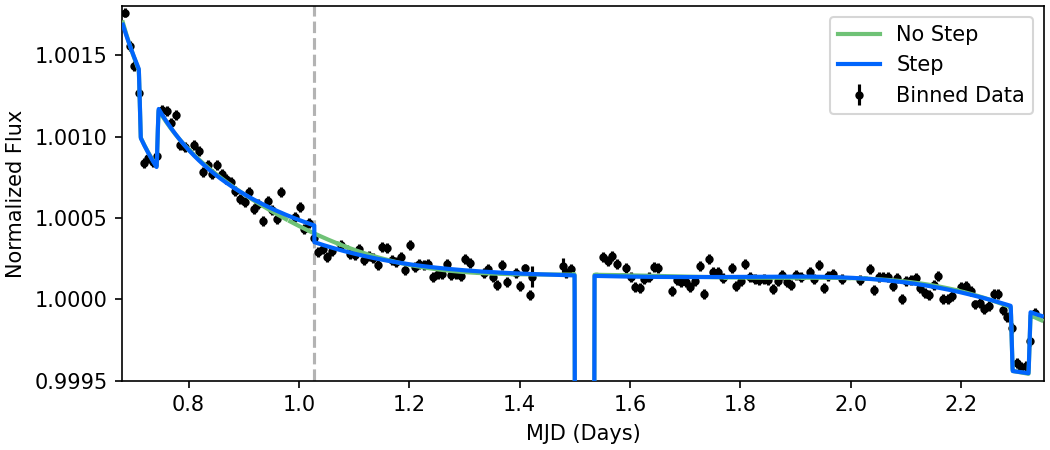}
    \includegraphics[width=0.6\linewidth]{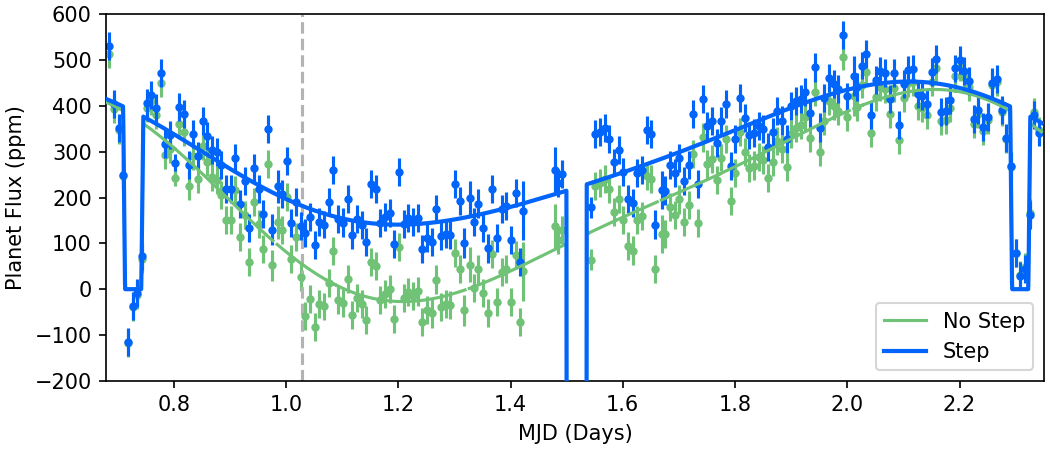}
    \caption{White light (5 -- 10.5 {\micron}) phase curves of GJ~1214b, with (blue) and without (green) fitting for a step function at 1.0278 days (gray dashed line).  In the top panel, the blue and green curves look nearly identical because the phase curve and systematic models are able to compensate for the lack of a step function.  The bottom panel highlights the significant difference between best-fit phase curve models.  The increase in nightside flux alters the planet's energy budget by decreasing the its inferred Bond albedo.  The blue light curve more closely matches the phase curve models shown in Figure \ref{fig:phasecurves}.}
    \label{fig:StepNoStep}
\end{figure}

\subsection{Phase-Resolved Emission Spectrum}

The dayside spectrum of GJ~1214b is unchanged between analyses.  Changes to the planet's emission spectrum at other orbital phases have a strong wavelength dependence, with longer wavelengths exhibiting systematically higher measured flux values relative to the analysis presented by \cite{Kempton2023}.  Figure \ref{fig:nightside} highlights these differences for the planet's nightside spectrum (where orbital phase = 0).  We note that the nightside spectrum longward of 10.5 {\microns} is unreliable and is only included for comparison purposes.

We also derive an updated set of heat redistribution efficiency values, $A_{obs}$.  Figure \ref{fig:Aobs} compares the two {\eureka} analyses to the values inferred by \cite{Kempton2023}.  In general, our revised points are lower longward of 7 {\microns}, which favors more efficient day-night heat recirculation.  Below 7 {\microns}, the nightside flux for all reductions is roughly consistent with zero and $A_{obs}$ trends toward unity.

\begin{figure}
    \centering
    \includegraphics[width=0.6\linewidth]{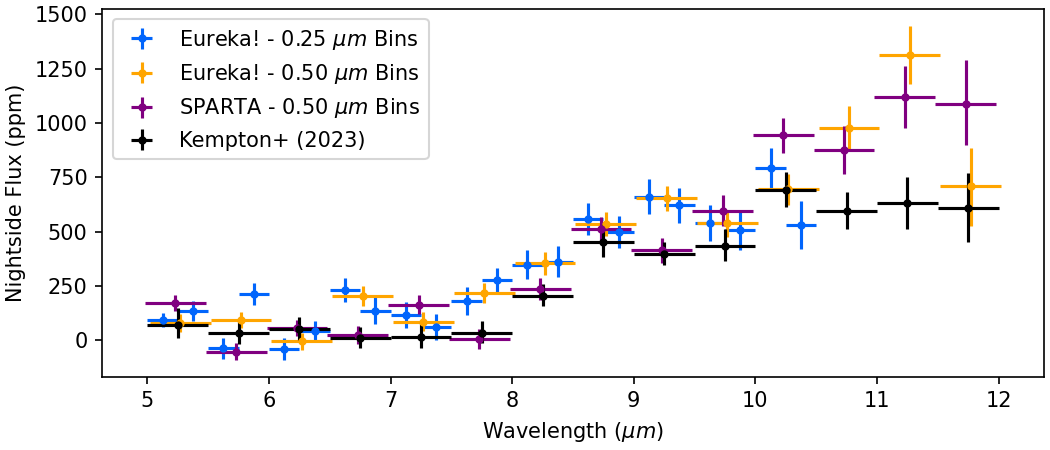}
    \caption{Revised nightside emission spectrum of GJ~1214b.  All three re-analyses (blue, orange, and purple) generally yield consistent results; however, the two \eureka~spectra are slightly higher than the SPARTA spectrum.  At longer wavelengths, the differences between the original nightside spectrum by \citet{Kempton2023} and the re-analyses becomes increasingly apparent.  The measured nightside emission from the blue and orange spectra more closely match the models shown in Figure \ref{fig:spectra}.}
    \label{fig:nightside}
\end{figure}

\begin{figure}
    \centering
    \includegraphics[width=0.6\linewidth]{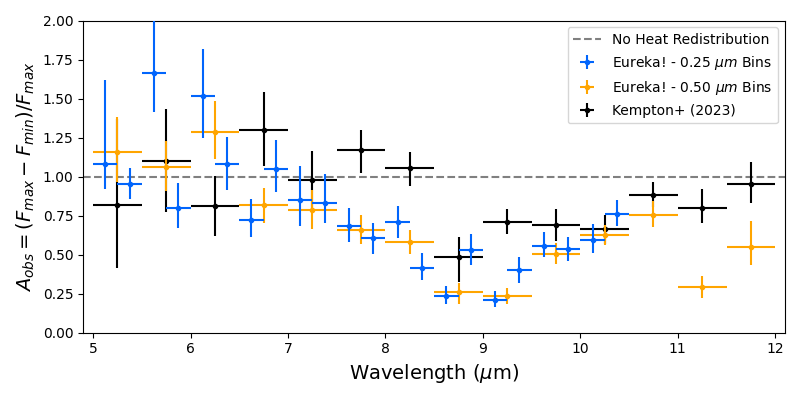}
    \caption{Revised phase curve amplitudes for GJ~1214b.  Both {\eureka} analyses (blue and orange points) are consistent; however, the 0.25-{\micron} reduction shows significant scatter between adjacent bins.
    Compared to those reported by \citet{Kempton2023} (black points), the revised $A_{obs}$ values are generally lower.  This is consistent with the warmer nightside flux seen in Figure \ref{fig:StepNoStep} when fitting for the step function at 1.0278 days.
    }
    \label{fig:Aobs}
\end{figure}

\clearpage
\newpage

\end{appendix}
\bibliographystyle{aasjournal}

\end{document}